# Conceptual Design of PID Detectors for the EicC Spectrometer[A]

Xin Li[a*], Yuxiang Zhao[a], Yutie Liang[a], Xu Sun[a]

*a Institute of Modern Physics, Chinese Academy of Sciences, 509 Nanchang Rd., Lanzhou 730000, China*

A R T I C L E I N F O

A B S T R A C T

The Electron-ion collider in China (EicC) is a proposed future electron-ion collider designed to achieve a high luminosity exceeding $2.0 \times 10^{33}$ cm$^{-2}$·s$^{-1}$, with a center-of-mass energy ranging from 15 to 20 GeV. Excellent particle identification (PID) with extensive momentum coverage is essential for investigating exclusive and semi-inclusive processes, as well as enabling precise 3D imaging of the nucleon structure in the EicC experiment. To meet its PID requirement, the EicC Collaboration has proposed the conceptual design of various Cherenkov detectors, including the DIRC in the barrel region and the RICH in the endcap region. It also involving the TOF detector for PID in the low momentum region. The GEANT4 simulation, which integrate advanced optical transmission models and image reconstruction algorithms, have been conducted to study and optimize the performance of these detectors.

## 1. Introduction

Investigating the origin of nucleon spin and mass within the framework of Quantum Chromodynamics (QCD) is crucial for understanding the fundamental structure of visible matter and the strong force that governs its interactions. To address these open questions, the Electron-Ion Collider in China (EicC) has been proposed as a dedicated facility for advanced research [1] based on the High Intensity Heavy-Ion Accelerator Facility (HIAF), which is currently under construction in Huizhou, China. The EicC experiment will feature highly polarized electrons (~80% polarization) colliding with heavy ions (~70% polarization) at a unique center-of-mass energy range of 15 to 20 GeV, with a luminosity of (2~4) × $10^{33}$ cm$^{-2}$s$^{-1}$. Additional experimental options include polarized deuterons and Helium-3, alongside unpolarized ion beams spanning from Carbon to Uranium. The conceptual design of the spectrometer is shown in Fig. 1 [2]. A solenoidal superconducting magnet (B = 1.5 T) is integrated around the tracking detector to enable precise charged particle momentum measurements. The momentum resolution is ~1% at 1 GeV in the barrel region ($-1 < \eta < 1.6$) and ~2% in the endcap regions ($-3 < \eta < -1$ and $1.6 < \eta < 3$), as shown in Fig. 2 [3].

One major challenge for the EicC spectrometer is to achieve large momentum PID capability up to 6 GeV/c in the barrel region and 15 GeV/c in the ion-endcap region for 3σ π/K separation. In the barrel region, drawing on BaBar [4] and EIC detector designs [5], a compact Detection of Internally Reflected Cherenkov lights (DIRC) detector has been implemented for PID. It is supplemented by a Low-Gain Avalanche Diode (LGAD) layer serving as a Time-of-Flight (TOF) detector close to DIRC. This setup enables precise measurement of the incident particle's hit position and time, significantly enhancing the Cherenkov angular reconstruction of DIRC. Simulation indicates that it can achieve 3σ π/K separation up to 6 GeV/c with an angular resolution around 1 mrad [6]. While in the ion-endcap region, to achieve the hadronic identification up to 15GeV/c, it requires a dual-radiator Ring Imaging Cherenkov detector (dRICH) utilizing aerogel and $C_2F_6$ gas as radiators. The TOF coupled with dRICH is expected to have an overall time resolution <50 ps (for a flight distance L = 1.8 m) for low momentum PID, covering the momentum range ≤2.1 GeV/c for 3σ π/K separation and ≤0.6 GeV/c for 3σ e/π separation. In the electron-endcap region, a modular Ring Imaging Cherenkov detector (mRICH) with an angular resolution ≤ 1.5mrad is a suitable option for the PID up to 4 GeV/c.

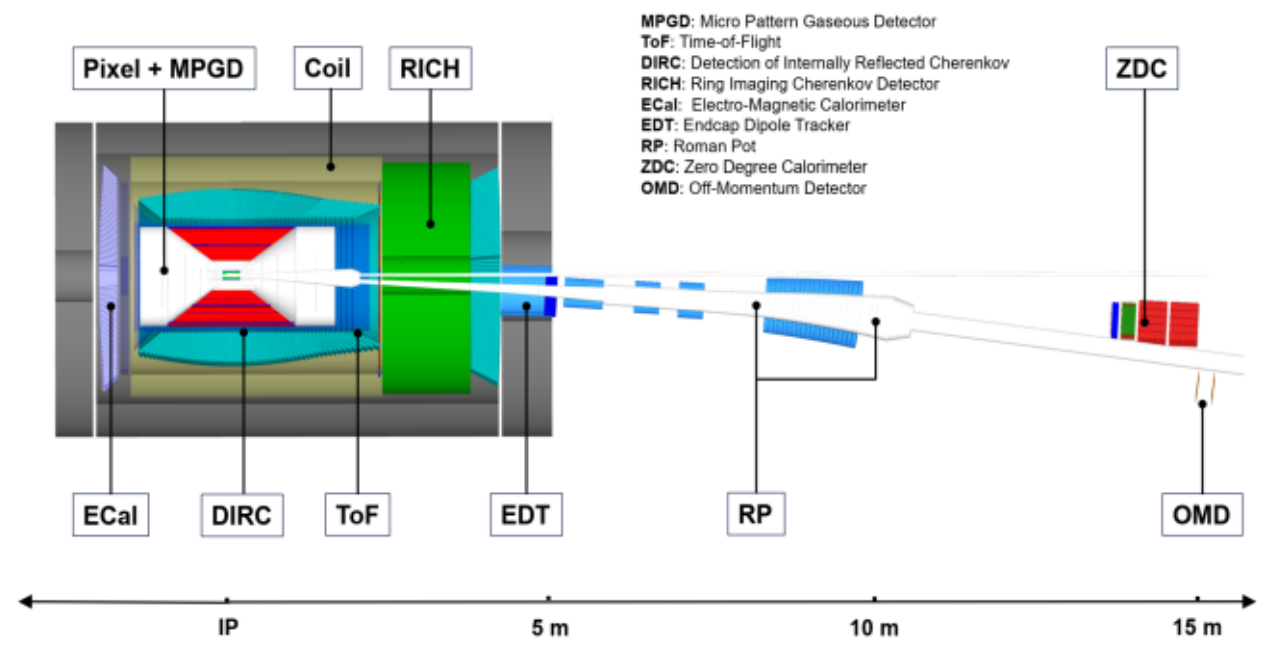

**Fig. 1.** Conceptual design of the EicC spectrometer.

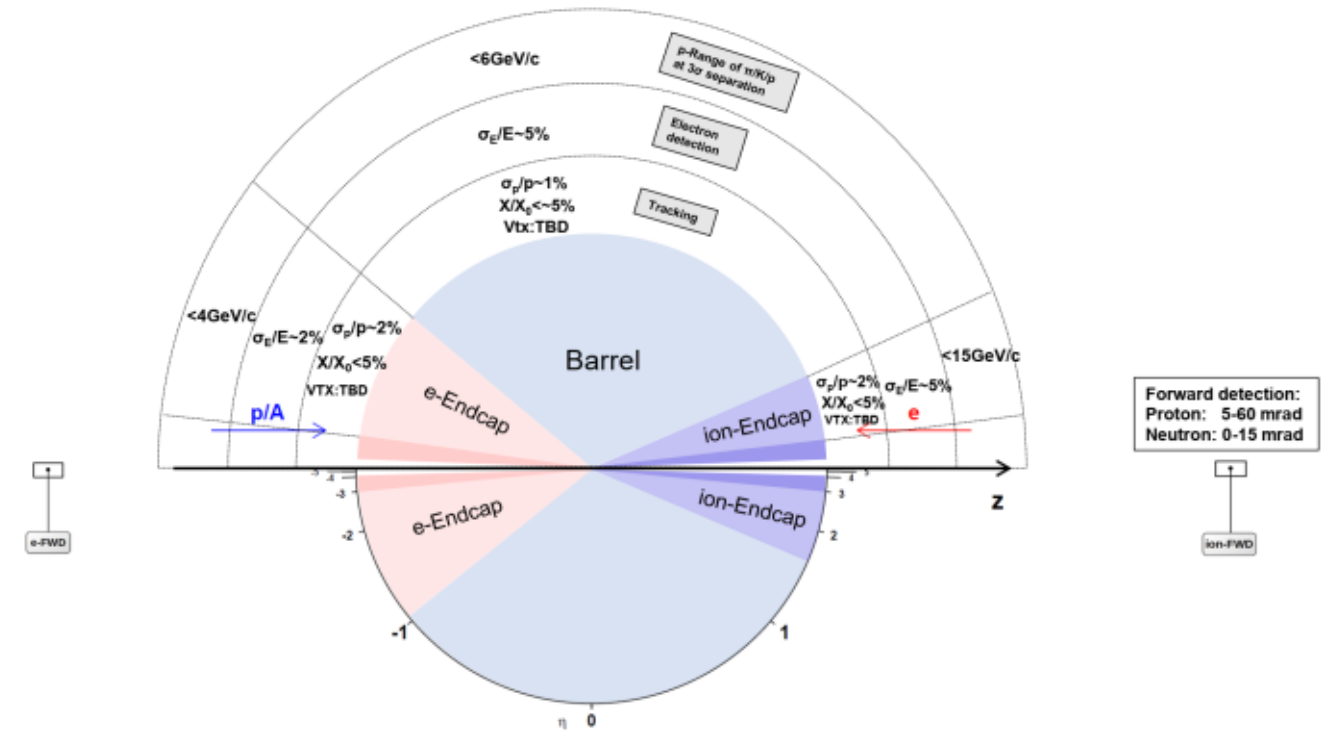

**Fig. 2.** EicC detector design specifications

[A] Supported by the National Key R&D Program of China, No. 2023YFA1606903, and National Natural Science Foundation of China, No. 12375185
* Corresponding author.
*E-mail address:* lixin@impcas.ac.cn (Xin Li).

## 2. PID detectors for the EicC spectrometer

In the EicC spectrometer, the PID system primarily consists of four detectors: DIRC, dRICH, mRICH, and ToF. Each Cherenkov detector employs different radiators to cover different momentum ranges. Table 1 lists the refractive indices of different radiators and their Cherenkov threshold momenta for different charged particles. The conceptual design and simulation results of these PID detectors will be presented in the following sections.

**Table 1** The refractive indices and Cherenkov threshold momenta of radiators

| Radiator | Refractive index | Threshold (GeV/c) | | | |
|---|---|---|---|---|---|
| | | e | π | k | P |
| Fused Silica(DIRC) | 1.473 | 0.00047 | 0.13 | 0.46 | 0.87 |
| Aerogel(mRICH) | 1.03 | 0.00213 | 0.58 | 2.06 | 3.92 |
| Aerogel(dRICH) | 1.02 | 0.00254 | 0.69 | 2.46 | 4.67 |
| $C_2F_6$ (dRICH) | 1.00080 | 0.0128 | 3.49 | 12.34 | 23.45 |

### *2.1. DIRC in the barrel region*

As a fused silica (n=1.47) Cherenkov detector, DIRC has compact structure, fast response, and large momentum coverage [7]. It not only saves valuable space in the barrel region and effectively reduces the cost of the spectrometer, but also meets the PID requirement in the high luminosity and radiation environment, making DIRC a preferred solution for the EicC experiment. The preliminary design of the barrel DIRC is shown in Fig. 3. Each DIRC tray consists of multiple rectangular fused silica radiators, an optical focus and expansion volume (EV), and MCP-PMT photosensor array. Twelve trays encircle the barrel region of the spectrometer, with the following design parameters:

-Quartz radiator bar: 15mm x 17mm x 3300mm;

-Expansion volume(EV): 208mm x 340mm x 300mm;

-MCP-PMT: Hamamatsu R10754 or N6021;

-Tray box size: 50mm x 320mm x 4000mm with 6 bar+EV;

-12 trays forms a barrel detector with a minimum radius R = 0.7m;

-Focusing mirror: spherical 3-layer lens (fused silica+N-LAK33b).

Based on the Geant4 simulation, the PID power of the barrel DIRC is estimated as shown in Fig. 4. With the angular resolution ~1mrad and the average number of collected photons >40, the DIRC's PID performance can meet the 3σ (s. d.) separation required by EicC. Within the measurable polar angle range, the 3σ π/K separation can achieve the momentum up to 6 GeV/c, while e/π separation can achieve the momentum up to 1.2 GeV/c.

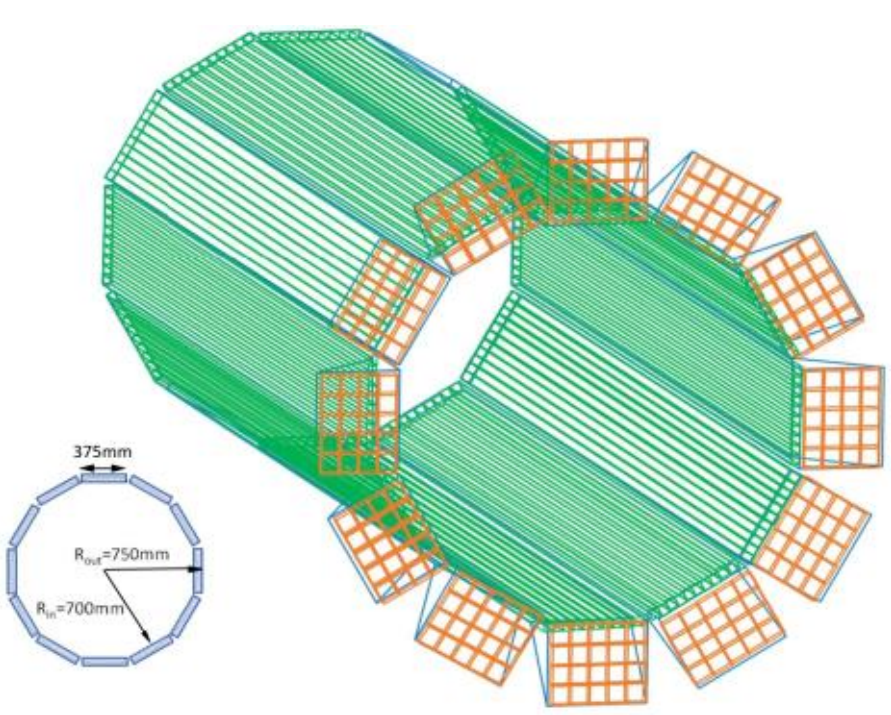

**Fig. 3.** The barrel DIRC conceptual design: each DIRC tray is consisted of 10 quartz radiator bars, a three-layer focusing lens + EV, and MCP-PMT array (orange) at one end (right); 12 trays form a barrel DIRC with a minimum radius R = 0.7 m (left).

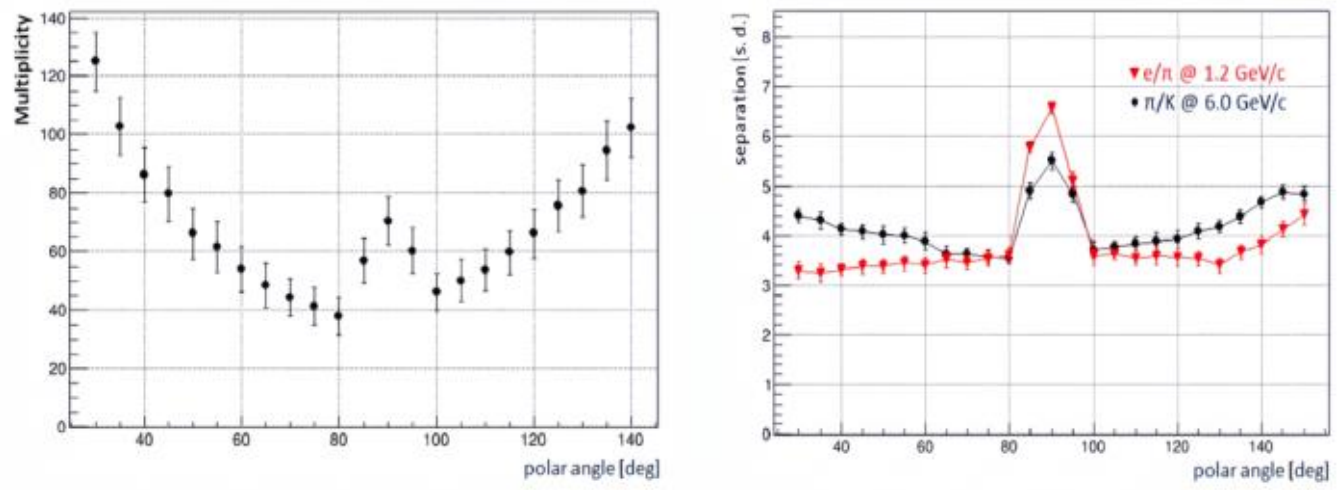

**Fig. 4.** The average number of collected photons (left) and the PID power (right) versus charged tracks' incidence angle in the barrel DIRC in GEANT4 simulation.

### *2.2. mRICH in the electron-endcap region*

In the mRICH, Cherenkov photons generated in the aerogel radiator are focused by a Fresnel lens and imaged on the PMT photosensor array. The focusing effect of the Fresnel lens can enhance the position resolution by limiting the wavelength range of transmitted light and suppressing Rayleigh scattering. The mRICH features a compact and flexible design, along with fast, powerful particle identification capabilities in a wide momentum range [8].

The mRICH in the electron endcap region consists of 64 aerogel modules (located at z=1080 ~ 1380 mm, radius=100~670mm) as shown in Fig. 5. The cross section of each module is 108 × 108 mm$^2$, with a thickness of 25 ~ 35 mm; The center of each module is positoned at z=-1230 mm and is tilted towards the collision center point; The associated Fresnel lens has a focal length L=76.2 mm (n=1.47, Edmund Optics). Fig. 6 shows its PID performance in GEANT4 simulation. The PID separation power of mRICH decreases as polar angle of incident particles increases: it can achieve 3σ π/K separation up to 9 GeV/c at best (when particle hit at the center of aerogel with the incident angle of 0 degree), and up to 8 GeV/c with the incident angle of 10 degree.

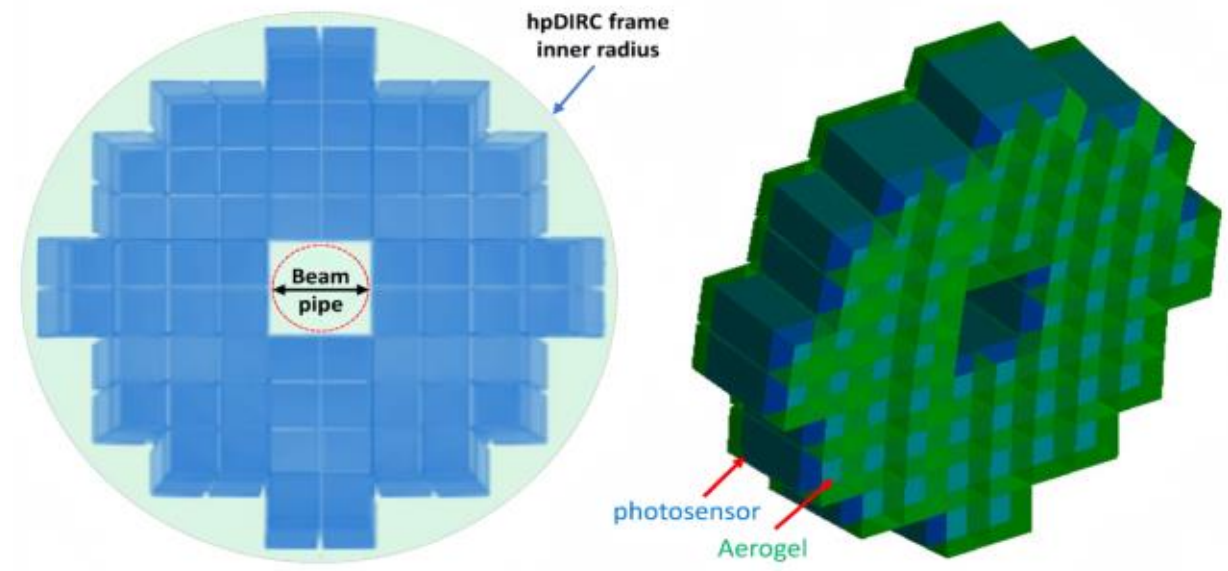

**Fig. 5.** The conceptual design of the mRICH in the electron endcap region.

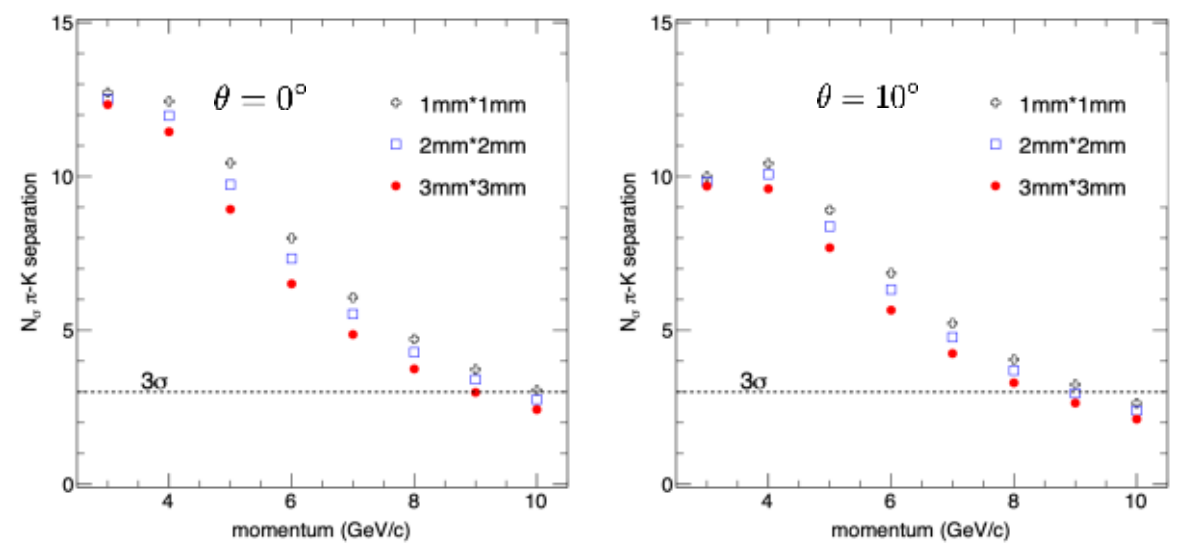

**Fig. 6.** The simulated PID performance of mRICH: it can achieve 3σ π/K separation up to 9 GeV/c at best with charged particles' incident angle of 0 degree (left), and 8 GeV/c with the incident angle of 10 degree (right). The millimeter-scale labels in figures represent different sizes of PMT pixels.

### 2.3. *dRICH in the ion-endcap region*

EicC PID in the ion-endcap region requires a continuous momentum coverage up to 15 GeV/c for hadron identification. Considering the redundancy for practical measurements, dRICH with dual radiators (aerogel + gas) is a suitable option. In the dRICH, charged particle passes through the aerogel and the gas sequencially, the induced Cherenkov radiation is focused by the spherical reflector and forms a halo image at the focal plane, which is subsequently readout by a photosensor array. Fig.7 shows the schematic of dRICH in the ion-endcap region based on the EIC design [9]. Its overall geometric parameters are as follows: length of 2160 mm, inner radius of 100 mm, outer radius of 1500 mm, and coverage angle ranging from 5 to 25 degrees. Its Cherenkov radiators include both aerogel (refractive index: n = 1.03/1.02 at 400 nm) and $C_2F_6$(n = 1.0008) . The thickness of each radiator is 40-50 mm for aerogel, and 1600-2000 mm for $C_2F_6$ gas.

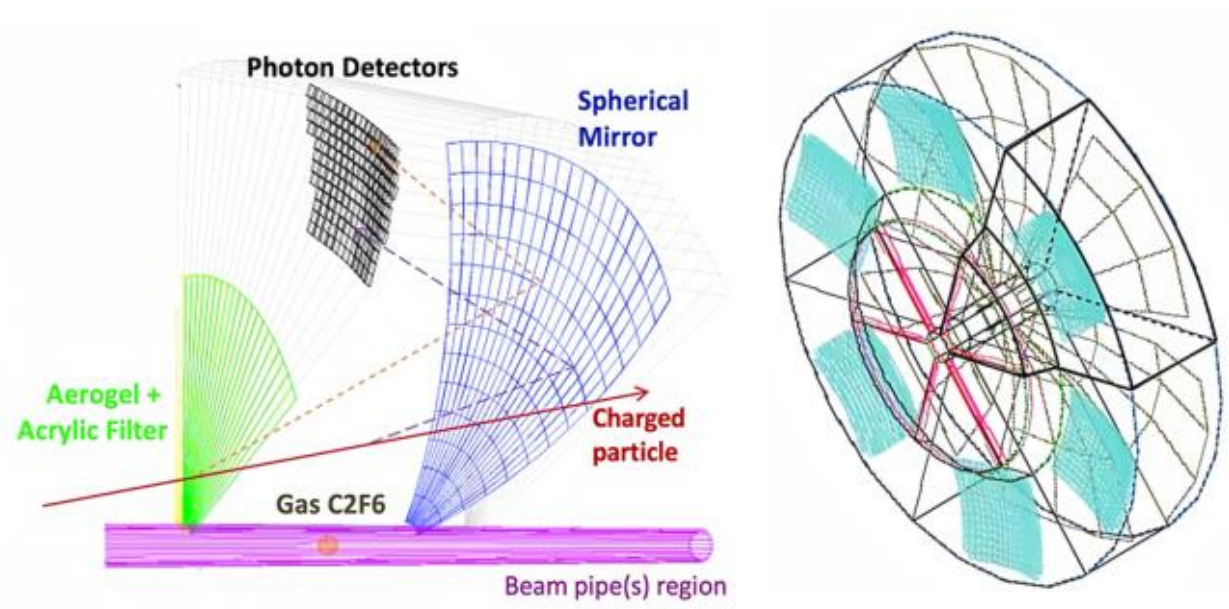

**Fig. 7.** Schematic concept design of ion-endcap dRICH detector

In GEANT4 simulation, assuming the reflectivity of the spherical mirror is 50% and the quantum efficiency of photosensors is 20%, approximately 60 photons are generated by the aerogel radiator per track. Considering the detection efficiency of the photosensor array, the actual measured number is 10~15pe. Meanwhile, approximately 200 photons are generated in the gas, with an actual measured number of 30~40pe. The minimum number of photoelectrons per track required for the dRICH detection is dependent on its image reconstruction efficiency, and typically 5 photoelectrons are sufficient. The GEANT4 simulation is still ongoing to further improve the photon collection efficiency by optimizing the optical geometry and image reconstruction method.

### 2.4. *LAGD TOF*

LGAD is a new type of position-sensitive semiconductor detector with high time resolution. Unlike other microstrip (or pixel) semiconductor detectors, LGAD has low gain and localized high electric field, which significantaly reduce the electron collection time [10]. It also has a compact multi-pixel structure and can provide high resolution (~30um) tracking information besides measuring ToF (20~30ps), making it the preferred TOF solution for EicC, providing hadronic identification below the threshold of Chernkov detectors, as well as low momentum electron identification. Due to its excellent position resolution, it will also effectively improve the accuracy of charged particles' tracking.

LGAD TOFs are installed in both barrel and endcap region of the EicC spectrometer, as shown in Fig. 8. The barrel TOF is located near the tracker system, the ion-endcap TOF is adjacent to the dRICH, and the electron-endcap TOF is right after the calorimeter. Its timing resolution is ~20ps per layer, and postition resolution is around 30 $\mu$m. The simulation of LAGD TOF is carried out using the minimum bias events, generated by PYTHIA in e (3.5 GeV) + p (20 GeV) collisions. Fig. 9 shows the simulated PID powers of TOF in two endcap regions. Both of them can achieve the 3σ hadronic identification up to 2.1GeV/c with a time resolution of 20ps.

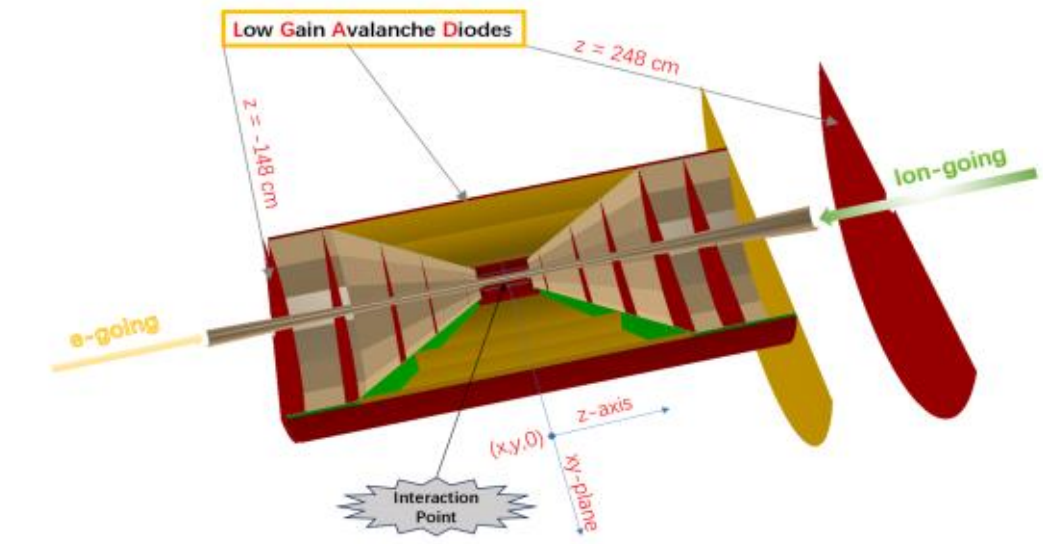

**Fig. 8.** The LGAD TOF in the EicC spectrometer.

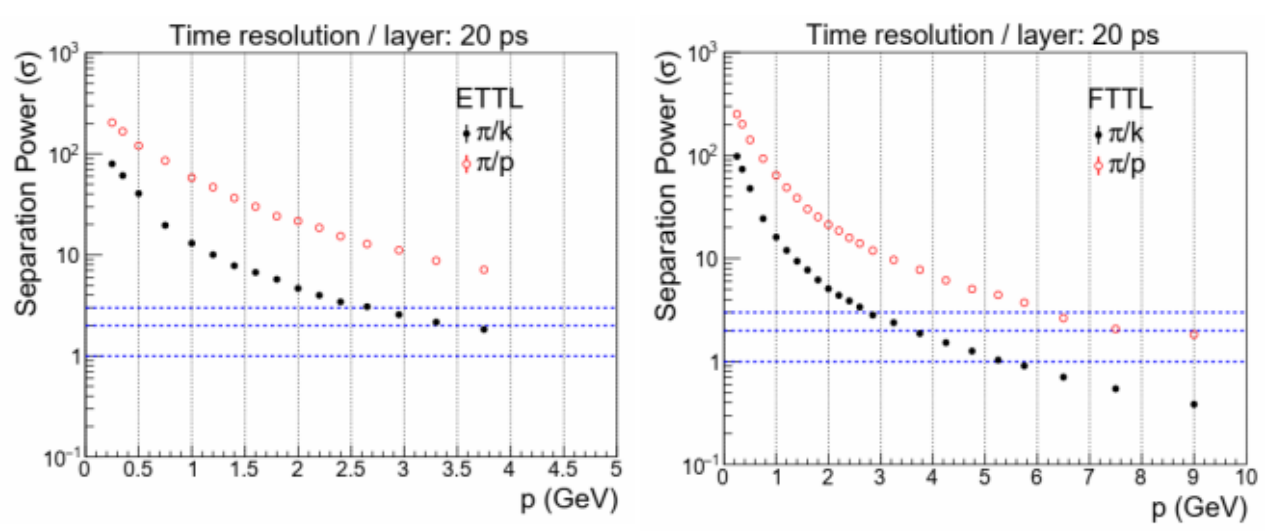

**Fig. 9.** The simulated PID powers of LGAD TOF in endcap regions (left: electron-endcap TOF, right: ion-endcap TOF).

## 3. Summary

Based on preliminary simulations and analysis, the proposed EicC PID system integrates barrel DIRC, endcap RICH, and LGAD TOF detectors. Their expected momentum coverage for 3σ π/K separation is shown in Fig. 10.

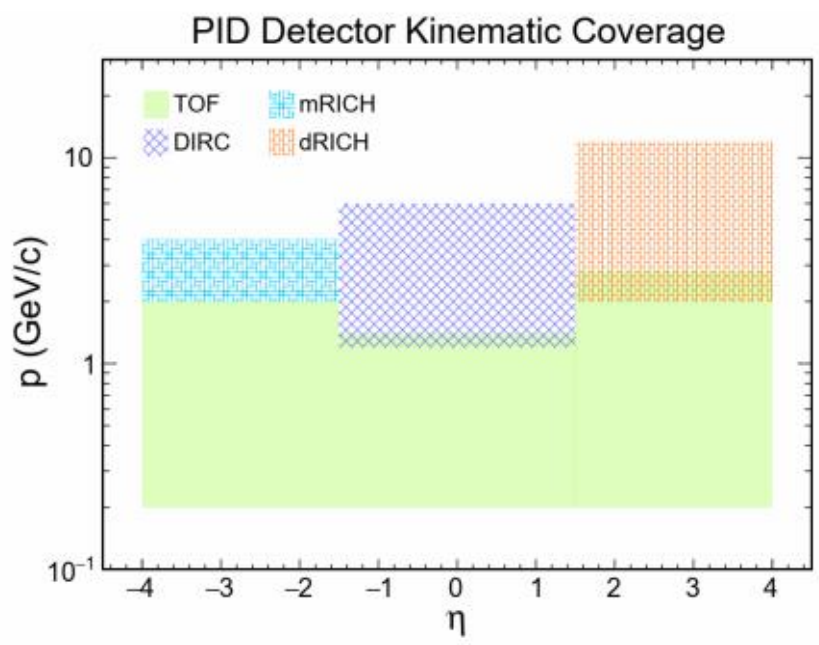

**Fig.10.** EicC PID detectors' momentum coverage for 3σ π/K separation.

This research work is supported by the National Key R&D Program of China, No. 2023YFA1606903, and National Natural Science Foundation of China, No. 12375185.